\documentclass[5p,twocolumn,sort&compress]{elsarticle}
\pdfoutput=1
\makeatletter					%
\def\ps@pprintTitle{				%
\let\@oddhead\@empty			%
\let\@evenhead\@empty			%
\def\@oddfoot{}					%
\let\@evenfoot\@oddfoot}			%
\makeatother					%
\usepackage{graphicx}
\usepackage{amsmath}
\usepackage{braket}
\usepackage{natbib}
\usepackage{hhline}
\usepackage[colorlinks=true,citecolor=blue,linkcolor=red,urlcolor=red]{hyperref}
\usepackage{comment}
\newcommand{\bq}{\begin{equation}}
\newcommand{\eq}{\end{equation}}
\renewcommand\Re{\operatorname{Re}}
\renewcommand\Im{\operatorname{Im}}
\newcommand\erfc{\operatorname{erfc}}

\begin{document}
\title{Decoherence due to gravitational time dilation: analysis of competing decoherence effects}
\author[DIP,INFN]{Matteo Carlesso}
    \ead{matteo.carlesso@ts.infn.it}

\author[DIP,INFN]{Angelo Bassi}
\ead{bassi@ts.infn.it}

\address[DIP]{Department of Physics, University of Trieste, Strada Costiera 11, 34151 Trieste, Italy}
\address[INFN]{Istituto Nazionale di Fisica Nucleare, Trieste Section, Via Valerio 2, 34127 Trieste, Italy}

\date{\today}
\begin{abstract}
Recently, a static gravitational field, such as that of the Earth, was proposed as a new source of decoherence \cite{Pikovski:2015aa}. We study the conditions under which it becomes the dominant  decoherence effect in typical interferometric experiments. The following competing sources are considered: spontaneous emission of light, absorption, scattering with the thermal photons and  collisions with the residual gas. We quantify all these effects. As we will see, current experiments are  off by several orders of magnitude. New ideas are needed in order to achieve the necessary requirements: having as large as system as possible, to increase gravitational decoherence,  cool it and isolated well enough to reduce thermal and collisional decoherence, and resolve very small distances. 

\end{abstract}
\maketitle
\section{Introduction}
Does quantum theory applies also to the macroscopic objects of everyday life? Quantum properties, the most relevant being quantum interference, are not visible at macroscopic scales, mainly due to decoherence~\cite{Joos:1985aa,Zurek:1991aa,Tegmark:1993aa,Joos:2003aa}: superpositions are washed away by the interaction with the surrounding environment, which makes it hard to detect them by interferometric experiments. The most common sources of decoherence are external random fields \cite{Unruh:1989aa}, residual gases \cite{Caldeira:1983aa,Hu:1992aa}, photons \cite{,Joos:1985aa,Jaynes:1963aa}. Recently a new source of decoherence was proposed by Pikovski et al.~\cite{Pikovski:2015aa}. This effect is due to gravitational time dilation of a system in a gravitational potential.
The authors consider a system consisting of a large number $N$ of harmonic oscillators in thermal equilibrium, in a gravitational potential. They show that the superposition in space of two center-of-mass wave packets decoheres when the two wave packets are centered in two positions, which have different gravitational potential, for example at two different heights on the Earth's surface.  

This effect is fascinating from the conceptual point of view. It is an example of entanglement between relative and center-of-mass degrees of freedom, mediated by the gravitational potential, and has already given rise to an intense debate \cite{Zych:2011aa,Bassi:2015ab,Gooding:2015aa,Zeh:2015ab,Pikovski:2015ab,Pikovski:2015ac,Diosi:2015aa,Margalit:2015aa,Bonder:2015ab,Adler:2016aa,Pang:2016aa,Pikovski:2016aa,Bonder:2016aa}. Here, we are interested in comparing this effect with other sources of decoherence, to understand under which conditions it can be detected, at least in principle.
\begin{figure}[b!]
\centerline{\includegraphics[width=1\linewidth]{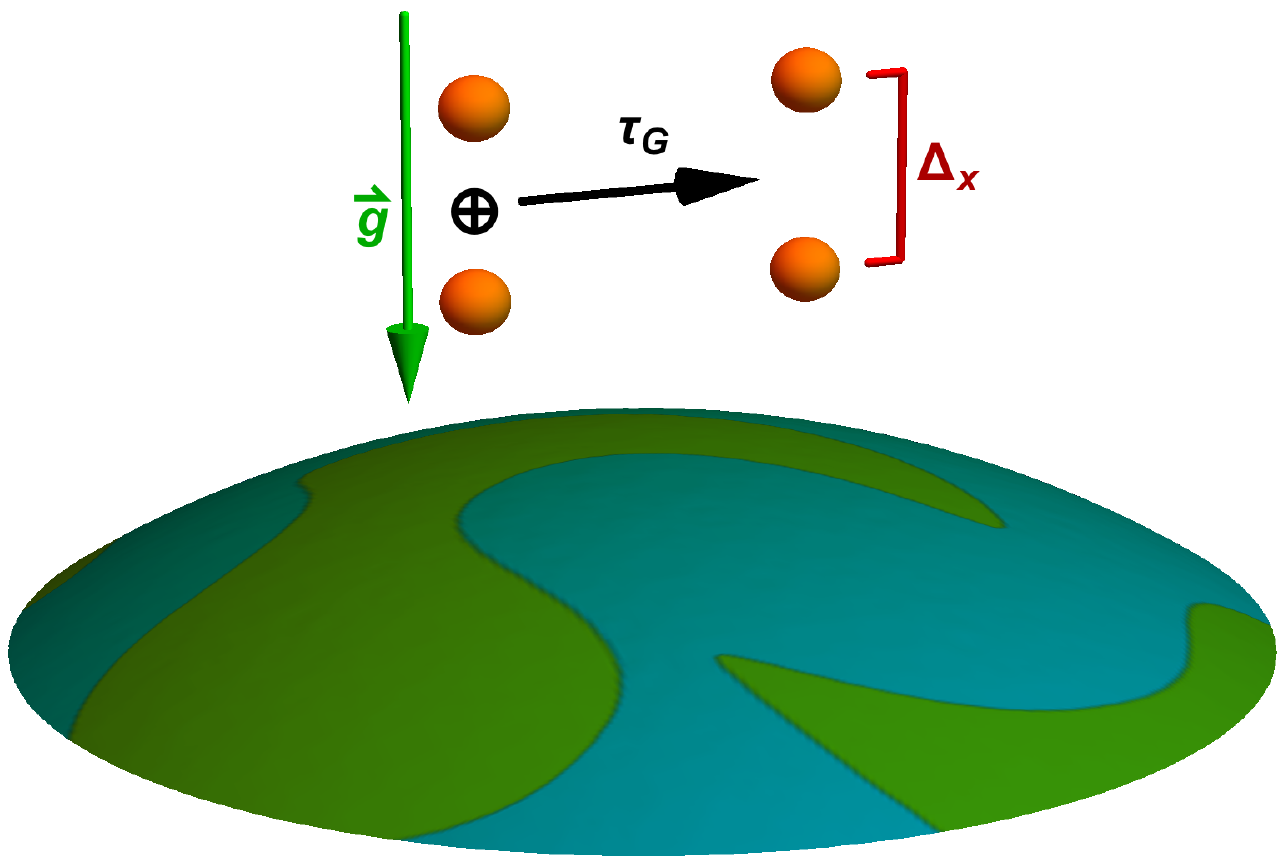}}
\caption{Graphical representation of the gravitational decoherence effect. The center of mass state of a system (for simplicity, we consider a sphere of radius $r$) is initially prepared in a vertical spatial superposition with separation $\Delta_x$. The coupling of the internal and center-of-mass degrees of freedom, mediated by the gravitational potential, decoheres the system. $\tau_\text{\tiny G}$ is the decoherence time.}
\end{figure}

The analysis done in \cite{Pikovski:2015aa}, which relies on the comparison between gravitational decoherence against  decoherence due to thermal emission, shows that, for low temperatures and small superposition distances, the former is dominant. However, as pointed-out in \cite{Adler:2016aa}, in typical interferometric experiments  also collisional decoherence with the residual gas should be taken into account, although the gas can be very diluted. 
We further explore this issue, from the quantitative point of view. First, we apply the Debye model for computing the heat capacity, instead of Einstein's model, which was used in~\cite{Pikovski:2015aa}; the former is more suitable for low  temperature regimes, the ideal scenario for detecting gravitational decoherence. We will show that Einstein's model overestimates the effect. Second, we will compare the gravitational effect with collisional decoherence, as well as thermal emission and scattering. We will see that there exist conditions for which gravity becomes the main decoherence mechanism, however these conditions are not easy to reach with foreseeable technology.

\section{Gravitational decoherence: comparison of Einstein and Debye model}
\label{sec:gravitation}
The decoherence time $\tau_\text{\tiny G}$ due gravitational decoherence on the Earth, written in terms of the heat capacity $C_\text{\tiny V}$ of the system, is \cite{Pikovski:2015aa}:
\bq
\tau_\text{\tiny G}=\dfrac{\sqrt{2}\hbar c^2}{\sqrt{K_\text{\tiny B}C_\text{\tiny V}}g T \Delta_x},
\eq
where $\hbar$ is the reduced Planck constant, $c$ the speed of light, $K_\text{\tiny B}$ the Boltzmann constant, $g$ the gravitational acceleration, $T$ the equilibrium temperature of the system, and $\Delta_x$ the vertical distance of a superposition of  center-of-mass states. As for the system, we consider a spherical crystal of radius $r$.

One crucial issue is how to model the heat capacity. A first simple model is provided by Einstein, who describes a crystal as made of independent harmonic oscillators, all having the same frequency \cite{Muller-Kirsten:2013aa}. The associated heat capacity gives an accurate description  for high temperatures. In this limit ($T>T_\text{\tiny D}$, where $T_\text{\tiny D}$ is the Debye temperature of the crystal) it reduces to the well known classical value $C^{\text{\tiny{CL}}}_\text{\tiny V}=N K_\text{\tiny B}$, where $N=3N_{\text{\tiny m}}$ is the number of degrees of freedom of the crystal and $N_{\text{\tiny m}}$ the number of molecules~\cite{Muller-Kirsten:2013aa}. This is the expression considered in \cite{Pikovski:2015aa}, which leads to the following formula for $\tau_\text{\tiny G}$:
\bq\label{tau_D_E}
\tau_\text{\tiny G}^\text{\tiny E}=\dfrac{\sqrt{2}\hbar c^2}{\sqrt{N}gK_\text{\tiny B} T \Delta_x}.
\eq
For high temperatures,  Einstein's model is a good approximation, however for low temperatures predictions deviate from the experimental data~\cite{Nash:1972aa}.  

A better model at low temperatures is provided by Debye. It assumes the crystal as made of independent harmonic oscillators distributed according to the Bose-Einstein statistics \cite{Mandl:1988aa}. The Debye heat capacity, in the limit $T\ll T_\text{\tiny D}$ of low temperatures, is \cite{Muller-Kirsten:2013aa} $C_\text{\tiny V}=4\pi^4/5\ N K_\text{\tiny B} (T/T_\text{\tiny D})^3$, which yields to the following expression for $\tau_\text{\tiny G}$:
\bq\label{tau_D_D_1}
\tau_\text{\tiny G}^\text{\tiny D}=\dfrac{1}{\pi^2}\sqrt{\dfrac{5}{2N}}\dfrac{\hbar c^2 T_\text{\tiny D}^{3/2}}{g K_\text{\tiny B} T^{5/2}\Delta_x}.
\eq
This is the gravitational decoherence time we will use for our (low temperature) analysis \cite{PrivatePikovski}. 
\begin{figure}
  \centering
  \setbox1=\hbox{\includegraphics[width=1\linewidth]{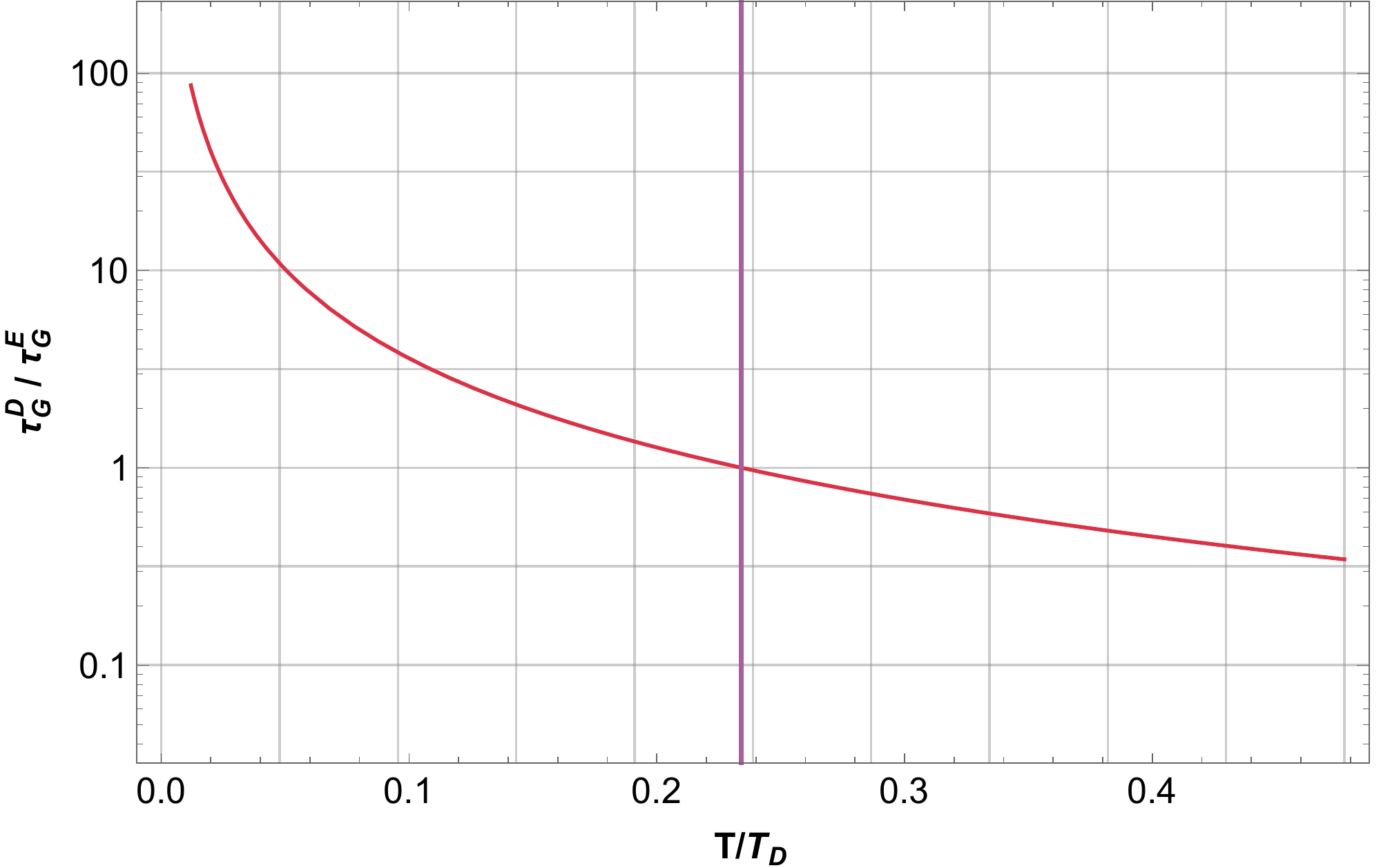}}
  \includegraphics[width=1\linewidth]{time_ratio.pdf}\llap{\raisebox{3cm}{\includegraphics[width=0.45\linewidth]{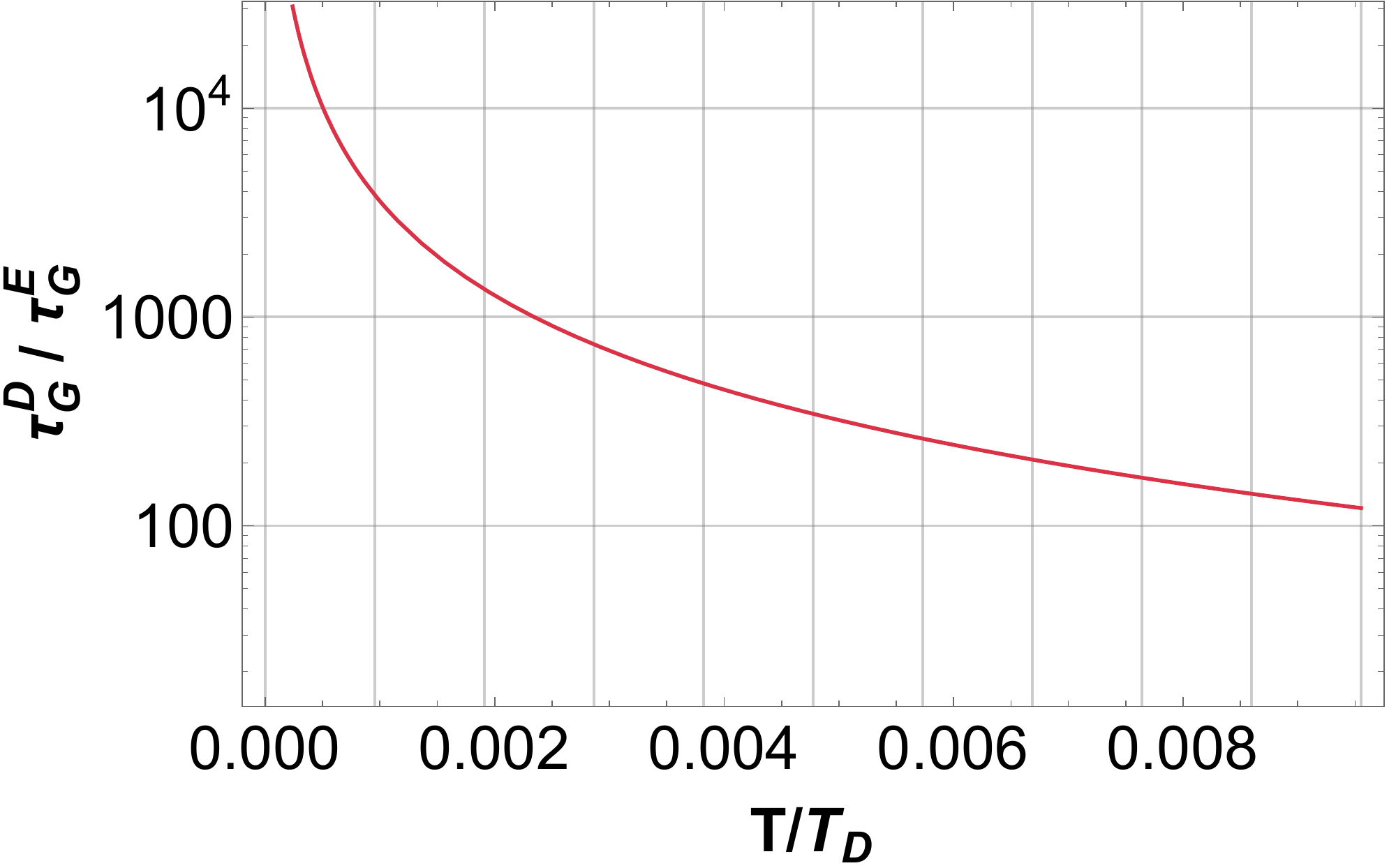}}} \\
\caption{Gravitational decoherence times as given by the Einstein ($\tau_\text{\tiny G}^{\text{\tiny E}}$) and Debye ($\tau_\text{\tiny G}^{\text{\tiny D}}$) models.  The plot shows the ratio  $\tau_\text{\tiny G}^\text{\tiny D}/\tau_\text{\tiny G}^{\text{\tiny E}}$ as a function of $T/T_\text{\tiny D}$. This ratio is independent from the superposition distance $\Delta_x$ and from the specifications of the system. The vertical purple line shows when $\tau_\text{\tiny G}^{\text{\tiny D}}$ and $\tau_\text{\tiny G}^{\text{\tiny E}}$ are equal.}
\label{plot_low_confronto}
\end{figure}

Figure \ref{plot_low_confronto} compares the two expressions for the gravitational decoherence time. As $\tau_\text{\tiny G}^\text{\tiny D}/\tau_\text{\tiny G}^{\text{\tiny E}}$ depends only on the ratio $T/T_\text{\tiny D}$, it is  independent from the specifications of the system, at least within the limits of validity of the models here considered. For temperatures smaller than $T_{\text{\tiny eq}}\sim 0.2\cdot T_\text{\tiny D}$, Einstein's model overestimates the gravitational decoherence effect, while it underestimates it for $T > T_{\text{\tiny eq}}$. Therefore, conceiving an experiment at very low temperatures to detect this gravitational effect is harder than originally estimated. 
As an explicit example, a sphere of radius $r =1\,\mu$m made of sapphire ($T_D = 1047$\,K) containing  $N_m\sim 10^{11}$ molecules, at an equilibrium temperature $T=1.0$\,K and delocalized over a distance $\Delta_x=10^{-3}\,$m (as considered in~\cite{Pikovski:2015aa}, see their Fig.~3), has a gravitational decoherence time $\tau_\text{\tiny G}^\text{\tiny D}\sim 6.9\times10^5$\,s according to Eq.~\eqref{tau_D_D_1}, which is three orders of magnitude longer than $\tau_\text{\tiny G}^{\text{\tiny E}}\sim1.8\times10^2$\,s, as given by Eq.~\eqref{tau_D_E}. 
\section{Competing effects}
\label{sec:compet}
To be visible, gravitational decoherence must be stronger than the other competing decoherence sources. We consider the two most common sources in experiments involving quantum superpositions of material systems: thermal and collisional decoherence. In both cases, the decay of the off-diagonal elements of the density matrix   is well described by the following expression~\cite{Schlosshauer:2007aa}:
\bq
\rho(x,y,t)=\rho(x,y,0)e^{-t/\tau_{\text{\tiny TC}}},
\eq
where the decoherence time $\tau_{\text{\tiny TC}}$ takes into account both thermal and collisional decoherence. This expression works well for recoil-free collisions (infinite mass limit) and low pressures; both conditions are satisfied in typical experiments as those here considered. Again, we consider a spherical crystal of radius $r$ made of $N_{\text{\tiny m}}=4\pi r^3 n/3$ molecules, where $n$ is the number density of molecules. Two different limits for $\tau_{\text{\tiny TC}}$ are relevant \cite{Joos:1985aa,Breuer:2002aa,Joos:2003aa,Romero-Isart:2011aa}. In the long wavelength limit ($2\pi\Delta_x\ll\lambda_{\text{\tiny dB}}$), where $\Delta_x = |x-y|$ and $\lambda_{\text{\tiny dB}}$ is the de Broglie wavelength of the system, $\tau_{\text{\tiny TC}} = 1/\Lambda\Delta_x^2$ where $\Lambda$ is the localization parameter characterizing the decoherence mechanism. In the short wavelength limit ($2\pi\Delta_x\gg\lambda_{\text{\tiny dB}}$) instead, $\tau_{\text{\tiny TC}} = \gamma^{-1}$, where $\gamma$ is rate of events. A reasonable ansatz for $\tau_{\text{\tiny TC}}$,  connecting the two limiting cases, is: 
\bq\label{tauthcoll}
\tau_{\text{\tiny TC}}=\left(\sum_i\gamma_i\tanh(\Delta_x^2\Lambda_i/\gamma_i	)\right)^{-1},
\eq
where the sum runs over all decoherence mechanisms (thermal and collisional, in our case).

Thermal decoherence includes three processes; scattering with the thermal environmental photons \cite{Joos:1985aa,Schlosshauer:2007aa,Romero-Isart:2011aa}
\bq
\Lambda_{\text{\tiny scatt}}=\dfrac{8! 8 \xi(9)c r^6}{9\pi}\left(\dfrac{K_\text{\tiny B}T}{\hbar c}\right)^9\left(\Re\left[\dfrac{\epsilon-1}{\epsilon+2}	\right]\right)^2,
\eq
where $\xi$ is the Riemann zeta function and $\epsilon$ is the complex dielectric constant of the crystal; absorption of the same environmental photons, in which case $\Lambda_{\text{\tiny abs}}=\Lambda_{\text{\tiny em}}$ \cite{Romero-Isart:2011aa}, where $\Lambda_{\text{\tiny em}}$ is the localization parameter associated to the spontaneous emission of photons, the third thermal process.
For the last two processes, we found two different models in the literature. The first, most widely used model (which we will refer to as Model 1), describes the system as a homogeneous particle of radius $r$, without taking its internal structure into account, in which case we have \cite{Romero-Isart:2011aa,Pikovski:2014aa}:
\bq\label{Lambda_em}
\Lambda_{\text{\tiny em}}^{(1)}=\dfrac{16\pi^5cr^3}{189}\left(	\dfrac{K_\text{\tiny B}T}{\hbar c}\right)^6\Im\left[\dfrac{\epsilon-1}{\epsilon+2}	\right].
\eq
The second model (Model 2) takes also the internal structure of the system into account, and for this reason shows an explicit dependence on the heat capacity. In this case:
\begin{multline}\label{Lambda_em2}
\Lambda_{\text{\tiny em}}^{(2)}=\dfrac{4 cr^3}{\pi }\left(\dfrac{K_\text{\tiny B}T}{ \hbar c}\right)^6 \Im\left[\dfrac{\epsilon-1}{\epsilon+2}\right]\lambda^3\cdot\\
\cdot\left[	2(\lambda+1)(\lambda+8)+\lambda^{1/2}(\lambda^2+10\lambda+15)e^{\lambda/2}\erfc(\sqrt{\lambda/2})	\right]
\end{multline}
where $\lambda=C_\text{\tiny V}/K_\text{\tiny B}$, and $\erfc(z)=1-2/\sqrt{\pi}\int_0^z dt\ e^{-t^2}$ is the complementary error function. In the following analysis we use both expressions for $\Lambda_\text{\tiny em}$ and consequently for $\Lambda_\text{\tiny abs}$. 

The rate of events, for all thermal processes, is the same and is given by \cite{Joos:2003aa}:
\bq
\gamma_{\text{\tiny th}}=\frac{2}{\pi}\xi(3)c r^2\left(\dfrac{K_\text{\tiny B}T}{\hbar c}	\right)^{3}.
\eq

Beside decoherence due to thermal photons, one typically has to consider also decoherence due to collisions with the residual gas particles, whose localization parameter is given by:
\bq\label{Lambda_coll}
\Lambda_{\text{\tiny coll}}=\dfrac{8\sqrt{2\pi}\xi(3)}{3\xi(3/2)}\sqrt{m_{\text{\tiny gas}}}n_{\text{\tiny gas}}\dfrac{r^2}{\hbar^2}(K_\text{\tiny B}T)^{3/2},
\eq
where $n_{\text{\tiny gas}}$ is the number density of the gas, which can be related to the pressure $P$ and temperature $T$, under the assumption of a dilute gas, using the ideal gas law: $n_{\text{\tiny gas}}=P/(K_BT)$. The rate instead is given by:
\bq\label{gamma_coll}
\gamma_{\text{\tiny coll}}=16\sqrt{3}\xi(3/2)\dfrac{P r^2}{\sqrt{m_{\text{\tiny gas}}K_\text{\tiny B}T}}.
\eq
These are the four main effects one typically takes into account when devising an interferometric experiment aimed at detecting quantum interference, or the lack of. The Appendix contains further details about the decoherence formulas, in particular the explicit derivation of the expressions in Eqs.~\eqref{Lambda_em2}, \eqref{Lambda_coll} and~\eqref{gamma_coll}.

\section{Comparison of the effects}
We compare the strength of the different decoherence sources. In Table \ref{tab:A} we consider some of the best known interferometric experiments, either already performed, or at the stage of proposal. 
\begin{table}[t!]      
\centering
\begin{tabular}{|c|ccc|}
\hline
\begin{tabular}[x]{@{}c@{}}Interferometric experiment\\(parameters)\end{tabular}& \begin{tabular}[x]{@{}c@{}} $t_{\text{\tiny exp}}$\\(s)\end{tabular} & \begin{tabular}[x]{@{}c@{}} $\tau_\text{\tiny G}^{\text{\tiny D}}$\\(s)\end{tabular} & \begin{tabular}[x]{@{}c@{}} $\tau_{\text{\tiny TC}}$\\(s)\end{tabular}  \\
\hline
\hline
\begin{tabular}[x]{@{}c@{}}Atoms\\($r\sim100$\,pm, $\Delta_x\sim 54$\,cm)\end{tabular}& $10^{-5}$  & $10^{29}$ & $10^{3}$ \\
\hline
\begin{tabular}[x]{@{}c@{}}Fullerenes\\($r\sim500$\,pm, $\Delta_x\sim100$\,nm)\end{tabular}& $10^{-2}$  & $10^{6}$ & $10^{-1}$ \\
\hline
\begin{tabular}[x]{@{}c@{}}Micro-particles\\($r=1\,\mu$m, $\Delta_x\sim 500$\,nm)\end{tabular}& $\ $  & $10^{12}$ & $1$\\
\hline
\begin{tabular}[x]{@{}c@{}}Diamonds\\($r\sim500$\,nm, $\Delta_x\sim 10$\,pm)\end{tabular}& $10^{-13}$  & $10^8$ & $10^{2}$ \\
\hline
\begin{tabular}[x]{@{}c@{}}Macro-particles\\($r=2$\,cm, $\Delta_x\sim 1$\,nm)\end{tabular}& $\ $  & $10^{3}$ & $10^{-19}$ \\
\hline
\end{tabular}
\caption{Experiments' time $t_{\text{\tiny exp}}$, gravitational decoherence time $\tau_\text{\tiny G}^{\text{\tiny D}}$ as given by Eq.~\eqref{tau_D_D_1} and thermal+collisional decoherence time $\tau_{\text{\tiny TC}}$  for some interferometric experiments with: atoms~\cite{Kovachy:2015aa}, fullerenes~\cite{Arndt:1999aa}, micro-particles~\cite{Pino:2016aa}, diamonds~\cite{Belli:2016aa}, macro-pacrticles~\cite{Schnabel:2015aa}. In all cases, we grossly simplified the system by shaping it as a homogeneous sphere of radius $r$. For thermal decoherence, we considered Model 1 of Section \ref{sec:compet}. The radius $r$ and delocalization distance $\Delta_x$ are shown in the table. The other parameters of the experiments are as follows. For $^{87}\text{Rb}$ atoms: $\epsilon\sim0.3 +0.1 i$, $P=10^{-17}\,$mbar, $T=10^{-9}\,$K. For C$_{60}$
fullerenes: $\epsilon\sim 4.4+10^{-3}i$, $P=5\times 10^{-17}\,$mbar, environmental $T=300\,$K, fullerenes' $T=900\,$K. For micro-particles of Nb: $\epsilon\sim41 + 10^{-4} i$, $P=10^{-17}\,$mbar and $T=30\,$mK. For diamonds: $\epsilon\sim5.7 +10^{-4}i$, $P=10^{-17}\,$mbar and $T=300\,$K. For macro-particles of $\text{SiO}_2$: $\epsilon\sim3.9 +10^{-3}i$, $P=5\times 10^{-7}\,$mbar and $T=4\,$K. In the case of micro and macro-particles, the experimental time $t_{\text{\tiny exp}}$ is not reported since these are theoretical proposals. In the case of fullerenes, the temperature of the experiment is well above the Debye temperature of the system ($T_\text{\tiny D}=185$\,K). In this case one has to use the Einstein's model, which gives $\tau_\text{\tiny G}^{\text{\tiny E}}\sim10^{8}$\,s.} 
\label{tab:A}
\end{table}
 The first column shows the duration of the experiment, the second the gravitational decoherence time $\tau_\text{\tiny G}$, computed using the Debye model; the third column displays the combined  thermal+collisional decoherence time  $\tau_{\text{\tiny TC}}$, using Model 1 for thermal decoherence. For simplicity, since we are interested in an order-of-magnitude analysis, we model all material systems as a homogeneous spheres of suitable radius. The numbers show that in all experiments so far performed, decoherence times are much longer that the experimental time, which is the reason why quantum interference was detected. However, in all cases the gravitational decoherence time is several orders of magnitude longer than the thermal+collisional decoherence time. This means that the current setups have to be significantly improved, to be used as detectors of gravitational decoherence.

\begin{figure}[t!]
\centering
\includegraphics[width=1\linewidth]{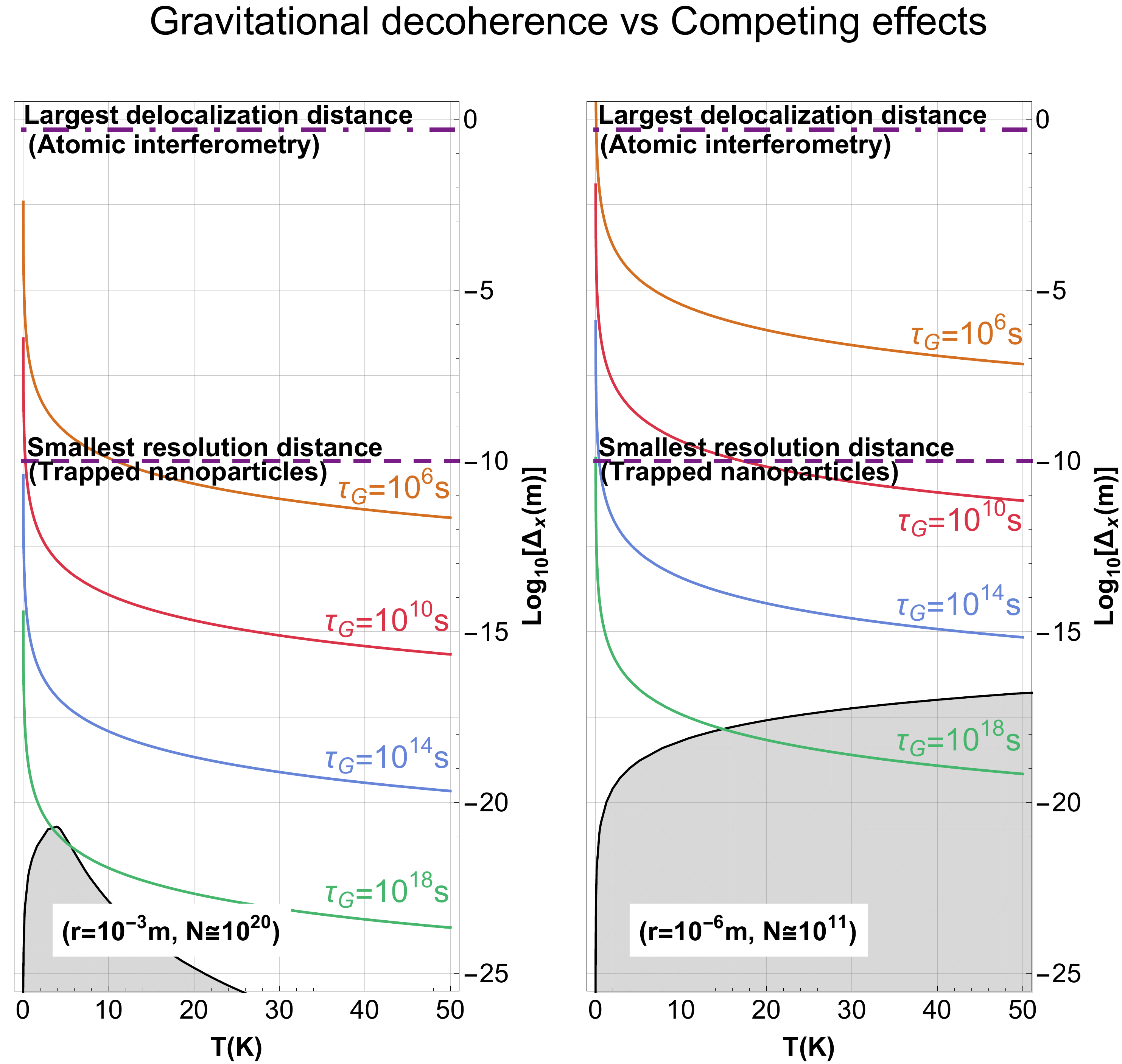}
\includegraphics[width=1\linewidth]{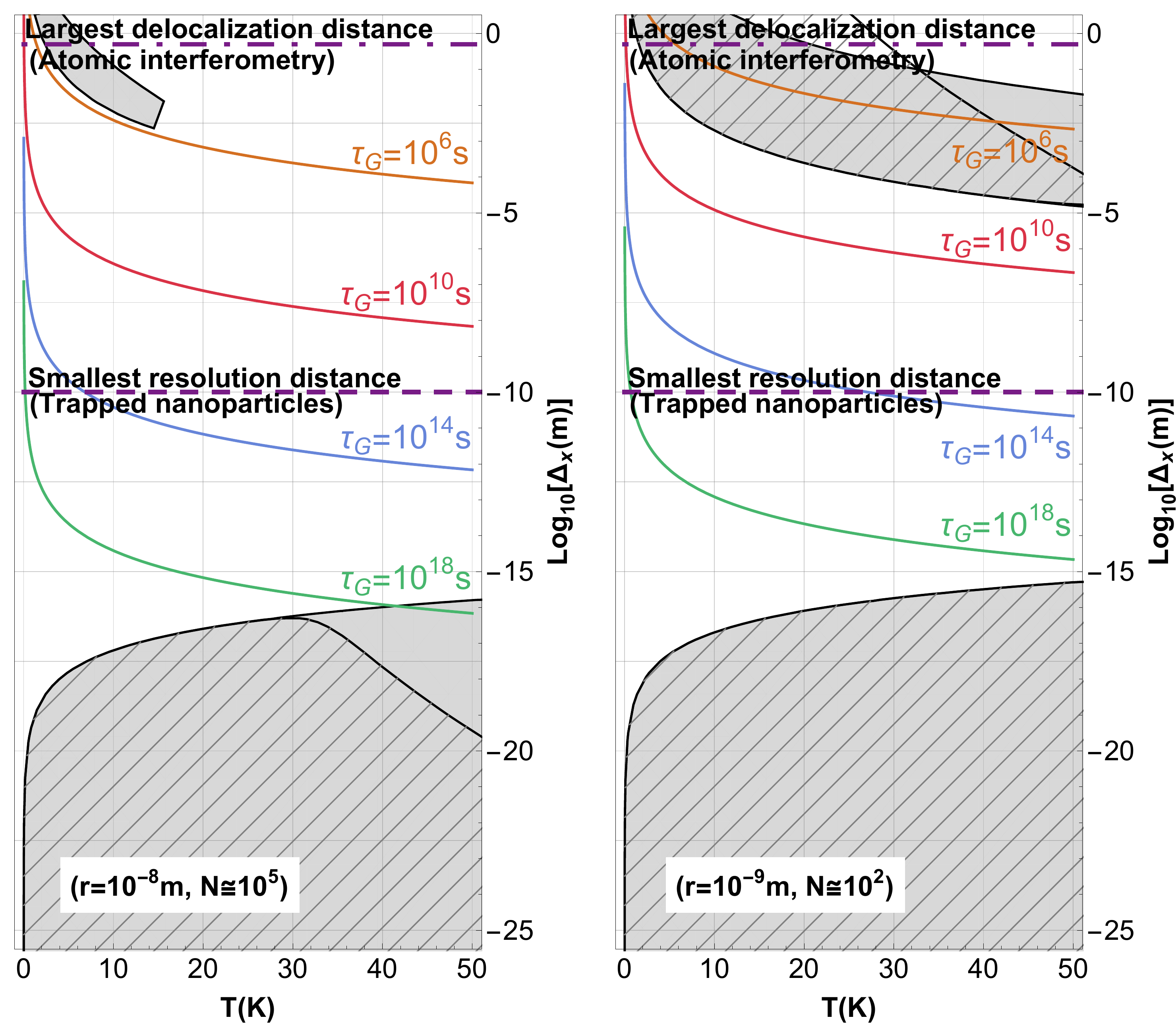}
\caption{Gravitational decoherence time $\tau_\text{\tiny G}$ vs.~thermal+collisional decoherence $\tau_{\text{\tiny TC}}$ as a function of the temperature $T$ and delocalization distance $\Delta_x$. We took a spherical crystal of sapphire (as considered in~\cite{Pikovski:2015aa}) of radius $r=10^{-3}$\,m (top left panel), $r=10^{-6}$\,m (top right panel), $r=10^{-8}$\,m (bottom left panel) and $r=10^{-9}$\,m (bottom right panel). The regions where gravitational decoherence is stronger, i.e. $\tau_\text{\tiny G} < \tau_{\text{\tiny TC}}$, are colored in grey. We considered  two models for thermal emission used in the literature, as discussed in Section \ref{sec:gravitation}: the region filled with grey refers to Model 1, that marked with diagonal grey lines to Model 2. We show also the maximum delocalization distance $\Delta_x$ currently achievable ($ 54$\,cm for atom interferometry \cite{Kovachy:2015aa}, purple dot-dashed line) and the minimum resolvable distance ($\sim 10^{-10}$\,m \cite{Gieseler:2012aa} in purple dashed line). The colored lines (orange, red, blue and green) show some numerical value of the gravitational decoherence times, as given by Eq.~\eqref{tau_D_D_1}.}
\label{fig:3}
\end{figure}

In Fig.~\ref{fig:3} we explore the parameter space temperature $T$ vs.~delocalization distance $\Delta_x$, to see for which values gravitational decoherence dominates the other effects. Following \cite{Pikovski:2015aa}, we choose sapphire, because of its low microwave emission at low temperatures \cite{Braginsky:1987aa,Lamb:1996aa,Pikovski:2015aa} and therefore for its suppressed thermal emission. The relevant parameters for sapphire are: $T_\text{\tiny D}=1047\,$K, $\rho=4.0\times 10^3\,$kg\,$\text{m}^{-3}$ and $\epsilon=10+10^{-9}i$. Again, we model the system as a homogeneous sphere of radius $r$. For the residual gas, we consider air, i.e.~a mixture of nitrogen $\text{N}_2$ at $78\%$ and oxygen $\text{O}_2$ at $22\%$ at the very low pressure of $P=10^{-17}\,\text{mbar}$, which is more o less the lowest pressure, which can be reached with existing technology \cite{Gabrielse:1990aa}.

The region where gravitational decoherence is stronger, i.e.~$\tau_\text{\tiny G}<\tau_{\text{\tiny TC}}$, is highlighted in grey. The are two types of regions. The one filled in grey corresponds to choosing Model 1 for thermal decoherence, while that marked with diagonal grey lines to Model 2 (see the discussion in Section \ref{sec:gravitation}).

As we can see, there are two disjoint grey regions. This is connected to the different dependence of the decoherence rate with respect to the delocalization distance $\Delta_x$. For gravitational decoherence, the dependence is linear, while for  thermal and collisional decoherence it is quadratic at short distances (long wavelength limit) and constant at large distances (short wavelength limit); see Eqs.~\eqref{tau_D_E} and~\eqref{tauthcoll}. More specifically, the lower grey region, present in all four panels, correspond to the regime where the long wavelength limit applies: $\tau_{\text{\tiny TC}}$ scales with $\Delta_x^2$, while $\tau_{\text{\tiny G}}$ scales with $\Delta_x$. The upper grey region, which appears only in the two bottom panels, correspond to  the short wavelength limit: in this case $\tau_{\text{\tiny TC}}$ is independent of $\Delta_x$, while $\tau_{\text{\tiny G}}$ still scales with $\Delta_x$. For this reason, there is a gap between the two grey regions in the bottom panels.

As Figure \ref{fig:3}  shows, there are regions where gravity is the dominant decoherence mechanisms. However, these regions are not easy to access experimentally, due to the quite extreme  conditions required, with respect to present-day technology. The first difficulty is related to the distances of has to resolve. For example, if we consider an optomechanical setup with a microparticle of radius $r=10^{-6}\,$m or larger, gravitational decoherence dominates over the other effects  if one is able to resolve distances $\Delta_x \sim 10^{-16}\,$m, which is smaller than the radius of the proton's charge. 

For smaller radii, as shown in the bottom panels ($r= 10^{-8}\,$m and $r= 10^{-9}\,$m), the numbers seem more favourable. But then one has to take another problem into account: the gravitational decoherence time ($10^{6}\sim10^{10}\,$s) is much longer than the typical time-scales of interferometric experiments ($\sim1\,$ms for matter-wave interferometry \cite{Eibenberger:2013aa} and $\sim 100\,$ms for optomechanics \cite{Bateman:2014aa}). 
\section{Conclusions}
To detect gravitational decoherence, interferometric experiments are the natural choice. One has to reduce standard decoherence sources (the two most important  being thermal and collisional), which is done by cooling the system and by placing the setup in vacuum. As shown in Table \ref{tab:A}, none of the typical interferometric experiments performed (or suggested) is capable of detecting gravitational decoherence, whose timescales are several orders of magnitude longer than the experiments' times.  New ideas are needed.

There are three difficulties to overcome. The first one is that for a system at thermal equilibrium at low temperatures, Einstein's model used in~\cite{Pikovski:2015aa} is not appropriate. A better model is that of Debye, which however weakens the predicted decoherence rate also by several orders of magnitude, depending on the temperature. This was discussed in Section \ref{sec:gravitation}.

A second difficulty is that, as shown in Fig.~\ref{fig:3}, the larger the size of the system, the stronger the competing effects (the smaller the grey area). This can be understood by comparing the decoherence time due to gravity with that due to the competing effects, for fixed $\Delta_x$: in the first case it scales with $\sqrt{N}\sim r^{3/2}$ and in the second case with $r^2$ or higher powers. In order for gravitational decoherence to be dominant, one has to consider very small delocalization distances, which cannot be easily resolved if the system is large ($\Delta_x < 10^{-16}$m for micrometric particles, or bigger ones).

The alternative is to use smaller systems, but in this case a third difficulty enter into play: the gravitational decoherence time simply becomes too long.

A possible way out is to consider the original effect presented in \cite{Zych:2011aa}, where only a few internal degrees of freedom (ideally, only two) are taken into account, in place of the many degrees of freedom considered in \cite{Pikovski:2015aa}. In this case the gravitational effect shows up as an oscillatory behavior in the visibility in an interferometric experiment, which can be distinguished more easily from standard decoherence. Also in this case the effect is very small, but potentially more likely to be detectable, e.g.~with cold atom experiments.
\section{Acknowledgements}
The authors wish to thank I.~Pikovski and H.~Ulbricht for many interesting and enjoyable discussions. They acknowledge financial support from the University of Trieste (FRA 2013), INFN and the John Templeton Foundation (grant N.~39530).
%


%
%
\section*{Appendix}
\label{App}
With reference to Section \ref{sec:compet}, here we derive  $\Lambda_{\text{\tiny em}}^{(2)}$ in Eq.~\eqref{Lambda_em2}, $\Lambda_{\text{\tiny coll}}$ in Eq.~\eqref{Lambda_coll} and $\gamma_\text{\tiny coll}$ in Eq.~\eqref{gamma_coll} in the low temperature limit, where the momentum distribution cannot be oversimplified by the Maxwell-Boltzmann law, as usually done in the literature. (All other quantities used in the main text take expressions, which are standard.)\\

The localization rate $\Lambda_{\text{\tiny em}}^{(2)}$ for thermal emission  is defined as follows \cite{Joos:1985aa,Breuer:2002aa,Pikovski:2015aa}:
\bq\label{def_Lambda_em_2}
\Lambda_{\text{\tiny em}}=c\int_0^{+\infty}dk\ k^2 N(k)g(k)\sigma_{\text{\tiny eff}}(k),
\eq
where $N(k)$ is the number of photons with wave vector $k$, $g(k)$ is the density of modes and $\sigma_{\text{\tiny eff}}(k)$ is the effective scattering cross section of the process. For black-body radiation \cite{Bohren:2008aa} the mode density is $g(k)=\pi^{-2}k^2$ (see~\cite{Pikovski:2014aa} for a further discussion).  As for the number of photons $N(k)$,  assuming a Planck distribution would not take into account  the internal structure of the crystal.  If we take it into account, we have~\cite{Frauendorf:1995aa,Hansen:1998aa,Hornberger:2004aa}:
\bq
N(k)=2 \exp\left[-\dfrac{\hbar c k}{K_\text{\tiny B} T}-\dfrac{K_\text{\tiny B}}{2C_\text{\tiny V}}\left(\dfrac{\hbar c k}{K_\text{\tiny B} T}	\right)^2	\right],
\eq
where the heat capacity $C_\text{\tiny V}$ conveys the information about the internal structure.

The cross section for spontaneous emission from a sphere of radius $r$, in the limit $k r\ll1$, is given by \cite{Frauendorf:1995aa,Bohren:2008aa}:
\bq
\sigma_{\text{\tiny eff}}(k)=4 \pi \Im[(\epsilon(k)-1)(\epsilon(k) +2)]k r^3,
\eq
where $\epsilon(k)$ is the complex dielectric constant of the crystal, which can be assumed not to change with $k$, ($\epsilon(k)\simeq \epsilon$). The two assumption ($kr\ll 1$ and $\epsilon(k)\simeq \epsilon$) are well justified; in fact the dominant contribution to the integral in Eq.~\eqref{def_Lambda_em_2} is given by small values of $k$. Eq.~\eqref{def_Lambda_em_2} then reduces to Eq.~\eqref{Lambda_em2}.

The general expression for localization rate $\Lambda_{\text{\tiny coll}}$ for collisional decoherence by a residual gas particles is \cite{Breuer:2002aa,Joos:2003aa,Romero-Isart:2011aa}
\begin{multline}
\Lambda_{\text{\tiny coll}}=\dfrac{2}{3}\dfrac{n_{\text{\tiny gas}}}{m_{\text{\tiny gas}}\hbar^2}\int_0^{+\infty}dp\ \nu(p)p^3\cdot\\
\cdot\int \dfrac{d\hat nd\hat n'}{4\pi}\sin^2(\theta/2)\left|F(p\hat n,p \hat n')	\right|^2
\label{Lambda_scatt_tutta}
\end{multline}
where $n_{\text{\tiny gas}}$ is the gas density, $\theta$ is the angle between the unitary vectors $\hat n$ and $\hat n'$, which define the directions of motion of the gas molecule before and after the scattering (with incoming momentum $p$), $F(p\hat n,p \hat n')$ is the scattering amplitude of the process and $\nu(p)$ describes the momentum distribution of the particles. For our low temperature analysis, we have to consider the Bose-Einstein distribution instead of the Maxwell-Boltzmann distribution, which is usually used in the literature~\cite{Breuer:2002aa,Joos:2003aa,Romero-Isart:2011aa}. We have:
\bq
\nu(p)=\sqrt{\dfrac{2}{\pi}}\dfrac{1}{\xi(3/2)(m_{\text{\tiny gas}}K_\text{\tiny B}T)^{3/2}}\dfrac{p^2}{e^{p^2/(2m_{\text{\tiny gas}}K_\text{\tiny B}T)}-1},
\eq
with $\int_0^{+\infty}dp\, \nu(p)=1$.  In the limit $\Delta_x\ll\lambda_{\text{\tiny dB}}^{\text{\tiny gas}}$ we can use the geometric cross section and Eq.~\eqref{Lambda_scatt_tutta} reduces to:
\bq
\Lambda_{\text{\tiny coll}}=\dfrac{\pi r^2 n_{\text{\tiny gas}}}{3\hbar^2 m_{\text{\tiny gas}}}\braket{p^3}_{\nu},
\eq
where $\braket{p^3}_{\nu}$ is computed with respect to the distribution $\nu(p)$:
\bq
\braket{p^3}_{\nu}=\int_0^{+\infty}dp\ \nu(p) p^3=8\sqrt{\dfrac{2}{\pi}}\dfrac{\xi(3)}{\xi(3/2)}(m_{\text{\tiny gas}}K_\text{\tiny B}T)^{3/2}.
\eq
We then obtain the expression in Eq.~\eqref{Lambda_coll} which, expressed in terms of the pressure $P$, becomes:
\bq
\Lambda_{\text{\tiny coll}}=\dfrac{8\sqrt{2\pi}\xi(3)}{3\xi(3/2)}P\dfrac{r^2}{\hbar^2}\sqrt{m_{\text{\tiny gas}}K_\text{\tiny B}T}.
\eq
Using the same distribution $\nu(p)$, we derive the rate $\gamma_\text{\tiny coll}$ from \cite{Romero-Isart:2011aa}:
\bq
\gamma_\text{\tiny coll}=\dfrac{16\pi\sqrt{2\pi}}{\sqrt{3}}\dfrac{Pr^2}{\braket{p}_{\nu}},
\eq
where
\bq
\braket{p}_{\nu}=\dfrac{\pi\sqrt{2\pi}}{3\xi(3/2)}\sqrt{m_\text{\tiny gas} K_\text{\tiny B}T}.
\eq
Combining these two expressions, we obtain \eqref{gamma_coll}.

\begin{thebibliography}{44}%
\makeatletter
\providecommand \@ifxundefined [1]{%
 \@ifx{#1\undefined}
}%
\providecommand \@ifnum [1]{%
 \ifnum #1\expandafter \@firstoftwo
 \else \expandafter \@secondoftwo
 \fi
}%
\providecommand \@ifx [1]{%
 \ifx #1\expandafter \@firstoftwo
 \else \expandafter \@secondoftwo
 \fi
}%
\providecommand \natexlab [1]{#1}%
\providecommand \enquote  [1]{``#1''}%
\providecommand \bibnamefont  [1]{#1}%
\providecommand \bibfnamefont [1]{#1}%
\providecommand \citenamefont [1]{#1}%
\providecommand \href@noop [0]{\@secondoftwo}%
\providecommand \href [0]{\begingroup \@sanitize@url \@href}%
\providecommand \@href[1]{\@@startlink{#1}\@@href}%
\providecommand \@@href[1]{\endgroup#1\@@endlink}%
\providecommand \@sanitize@url [0]{\catcode `\\12\catcode `\$12\catcode
  `\&12\catcode `\#12\catcode `\^12\catcode `\_12\catcode `\%12\relax}%
\providecommand \@@startlink[1]{}%
\providecommand \@@endlink[0]{}%
\providecommand \url  [0]{\begingroup\@sanitize@url \@url }%
\providecommand \@url [1]{\endgroup\@href {#1}{\urlprefix }}%
\providecommand \urlprefix  [0]{URL }%
\providecommand \Eprint [0]{\href }%
\providecommand \doibase [0]{http://dx.doi.org/}%
\providecommand \selectlanguage [0]{\@gobble}%
\providecommand \bibinfo  [0]{\@secondoftwo}%
\providecommand \bibfield  [0]{\@secondoftwo}%
\providecommand \translation [1]{[#1]}%
\providecommand \BibitemOpen [0]{}%
\providecommand \bibitemStop [0]{}%
\providecommand \bibitemNoStop [0]{.\EOS\space}%
\providecommand \EOS [0]{\spacefactor3000\relax}%
\providecommand \BibitemShut  [1]{\csname bibitem#1\endcsname}%
\let\auto@bib@innerbib\@empty
\bibitem [{\citenamefont {Pikovski}\ \emph
  {et~al.}(2015{\natexlab{a}})\citenamefont {Pikovski}, \citenamefont {Zych},
  \citenamefont {Costa},\ and\ \citenamefont {Brukner}}]{Pikovski:2015aa}%
  \BibitemOpen
  \bibfield  {author} {\bibinfo {author} {\bibfnamefont {I.}~\bibnamefont
  {Pikovski}}, \bibinfo {author} {\bibfnamefont {M.}~\bibnamefont {Zych}},
  \bibinfo {author} {\bibfnamefont {F.}~\bibnamefont {Costa}}, \ and\ \bibinfo
  {author} {\bibfnamefont {C.}~\bibnamefont {Brukner}},\ }\href
  {http://dx.doi.org/10.1038/nphys3366} {\bibfield  {journal} {\bibinfo
  {journal} {Nat.~Phys.}\ }\textbf {\bibinfo {volume} {11}},\ \bibinfo {pages}
  {668} (\bibinfo {year} {2015}{\natexlab{a}})}\BibitemShut {NoStop}%
\bibitem [{\citenamefont {Joos}\ and\ \citenamefont {Zeh}(1985)}]{Joos:1985aa}%
  \BibitemOpen
  \bibfield  {author} {\bibinfo {author} {\bibfnamefont {E.}~\bibnamefont
  {Joos}}\ and\ \bibinfo {author} {\bibfnamefont {H.~D.}\ \bibnamefont {Zeh}},\
  }\href {http://dx.doi.org/10.1007/BF01725541} {\bibfield  {journal} {\bibinfo
   {journal} {Z.~Phys.~B}\ }\textbf {\bibinfo {volume} {59}},\ \bibinfo {pages}
  {223} (\bibinfo {year} {1985})}\BibitemShut {NoStop}%
\bibitem [{\citenamefont {Zurek}(1991)}]{Zurek:1991aa}%
  \BibitemOpen
  \bibfield  {author} {\bibinfo {author} {\bibfnamefont {W.~H.}\ \bibnamefont
  {Zurek}},\ }\href {http://dx.doi.org/10.1063/1.881293} {\bibfield  {journal}
  {\bibinfo  {journal} {Phys.~Today}\ }\textbf {\bibinfo {volume} {44}},\
  \bibinfo {pages} {36} (\bibinfo {year} {1991})}\BibitemShut {NoStop}%
\bibitem [{\citenamefont {Tegmark}(1993)}]{Tegmark:1993aa}%
  \BibitemOpen
  \bibfield  {author} {\bibinfo {author} {\bibfnamefont {M.}~\bibnamefont
  {Tegmark}},\ }\href {\doibase 10.1007/BF00662807} {\bibfield  {journal}
  {\bibinfo  {journal} {Found.~Phys.~Lett.}\ }\textbf {\bibinfo {volume} {6}},\
  \bibinfo {pages} {571} (\bibinfo {year} {1993})}\BibitemShut {NoStop}%
\bibitem [{\citenamefont {Joos}\ \emph {et~al.}(2003)\citenamefont {Joos},
  \citenamefont {Zeh}, \citenamefont {Kiefer}, \citenamefont {Giulini},
  \citenamefont {Kupsch},\ and\ \citenamefont {Stamatescu}}]{Joos:2003aa}%
  \BibitemOpen
  \bibfield  {author} {\bibinfo {author} {\bibfnamefont {E.}~\bibnamefont
  {Joos}}, \bibinfo {author} {\bibfnamefont {H.~D.}\ \bibnamefont {Zeh}},
  \bibinfo {author} {\bibfnamefont {C.}~\bibnamefont {Kiefer}}, \bibinfo
  {author} {\bibfnamefont {D.~J.~W.}\ \bibnamefont {Giulini}}, \bibinfo
  {author} {\bibfnamefont {J.}~\bibnamefont {Kupsch}}, \ and\ \bibinfo {author}
  {\bibfnamefont {I.-O.}\ \bibnamefont {Stamatescu}},\ }\href@noop {} {\emph
  {\bibinfo {title} {{Decoherence and the Appearance of a Classical World in
  Quantum Theory}}}},\ \bibinfo {edition} {2nd}\ ed.\ (\bibinfo  {publisher}
  {Springer-Verlag Berlin Heidelberg},\ \bibinfo {year} {2003})\BibitemShut
  {NoStop}%
\bibitem [{\citenamefont {Unruh}\ and\ \citenamefont
  {Zurek}(1989)}]{Unruh:1989aa}%
  \BibitemOpen
  \bibfield  {author} {\bibinfo {author} {\bibfnamefont {W.~G.}\ \bibnamefont
  {Unruh}}\ and\ \bibinfo {author} {\bibfnamefont {W.~H.}\ \bibnamefont
  {Zurek}},\ }\href {\doibase 10.1103/PhysRevD.40.1071} {\bibfield  {journal}
  {\bibinfo  {journal} {Phys.~Rev.~D}\ }\textbf {\bibinfo {volume} {40}},\
  \bibinfo {pages} {1071} (\bibinfo {year} {1989})}\BibitemShut {NoStop}%
\bibitem [{\citenamefont {Caldeira}\ and\ \citenamefont
  {Leggett}(1983)}]{Caldeira:1983aa}%
  \BibitemOpen
  \bibfield  {author} {\bibinfo {author} {\bibfnamefont {A.~O.}\ \bibnamefont
  {Caldeira}}\ and\ \bibinfo {author} {\bibfnamefont {A.~J.}\ \bibnamefont
  {Leggett}},\ }\href {http://dx.doi.org/10.1016/0378-4371(83)90013-4}
  {\bibfield  {journal} {\bibinfo  {journal} {Phys.~A}\ }\textbf {\bibinfo
  {volume} {121}},\ \bibinfo {pages} {587 } (\bibinfo {year}
  {1983})}\BibitemShut {NoStop}%
\bibitem [{\citenamefont {Hu}\ \emph {et~al.}(1992)\citenamefont {Hu},
  \citenamefont {Paz},\ and\ \citenamefont {Zhang}}]{Hu:1992aa}%
  \BibitemOpen
  \bibfield  {author} {\bibinfo {author} {\bibfnamefont {B.~L.}\ \bibnamefont
  {Hu}}, \bibinfo {author} {\bibfnamefont {J.~P.}\ \bibnamefont {Paz}}, \ and\
  \bibinfo {author} {\bibfnamefont {Y.}~\bibnamefont {Zhang}},\ }\href
  {\doibase 10.1103/PhysRevD.45.2843} {\bibfield  {journal} {\bibinfo
  {journal} {Phys.~Rev.~D}\ }\textbf {\bibinfo {volume} {45}},\ \bibinfo
  {pages} {2843} (\bibinfo {year} {1992})}\BibitemShut {NoStop}%
\bibitem [{\citenamefont {Jaynes}\ and\ \citenamefont
  {Cummings}(1963)}]{Jaynes:1963aa}%
  \BibitemOpen
  \bibfield  {author} {\bibinfo {author} {\bibfnamefont {E.~T.}\ \bibnamefont
  {Jaynes}}\ and\ \bibinfo {author} {\bibfnamefont {F.~W.}\ \bibnamefont
  {Cummings}},\ }\href {\doibase 10.1109/PROC.1963.1664} {\bibfield  {journal}
  {\bibinfo  {journal} {Proc.~IEEE}\ }\textbf {\bibinfo {volume} {51}},\
  \bibinfo {pages} {89} (\bibinfo {year} {1963})}\BibitemShut {NoStop}%
\bibitem [{\citenamefont {Zych}\ \emph {et~al.}(2011)\citenamefont {Zych},
  \citenamefont {Costa}, \citenamefont {Pikovski},\ and\ \citenamefont
  {Brukner}}]{Zych:2011aa}%
  \BibitemOpen
  \bibfield  {author} {\bibinfo {author} {\bibfnamefont {M.}~\bibnamefont
  {Zych}}, \bibinfo {author} {\bibfnamefont {F.}~\bibnamefont {Costa}},
  \bibinfo {author} {\bibfnamefont {I.}~\bibnamefont {Pikovski}}, \ and\
  \bibinfo {author} {\bibfnamefont {C.}~\bibnamefont {Brukner}},\ }\href
  {http://dx.doi.org/10.1038/ncomms1498} {\bibfield  {journal} {\bibinfo
  {journal} {Nat.~Commun.}\ }\textbf {\bibinfo {volume} {2}},\ \bibinfo {pages}
  {505} (\bibinfo {year} {2011})}\BibitemShut {NoStop}%
  \bibitem [{\citenamefont {Bassi}(2015)}]{Bassi:2015ab}%
  \BibitemOpen
  \bibfield  {author} {\bibinfo {author} {\bibfnamefont {A.}~\bibnamefont
  {Bassi}},\ }\href {http://dx.doi.org/10.1038/nphys3390} {\bibfield  {journal}
  {\bibinfo  {journal} {Nat.~Phys.}\ }\textbf {\bibinfo {volume} {11}},\
  \bibinfo {pages} {626} (\bibinfo {year} {2015})}\BibitemShut {NoStop}%
\bibitem [{\citenamefont {Gooding}\ and\ \citenamefont
  {Unruh}(2015)}]{Gooding:2015aa}%
  \BibitemOpen
  \bibfield  {author} {\bibinfo {author} {\bibfnamefont {C.}~\bibnamefont
  {Gooding}}\ and\ \bibinfo {author} {\bibfnamefont {W.~G.}\ \bibnamefont
  {Unruh}},\ }\href {http://dx.doi.org/10.1007/s10701-015-9939-9} {\bibfield
  {journal} {\bibinfo  {journal} {Found.~Phys.}\ }\textbf {\bibinfo {volume}
  {45}},\ \bibinfo {pages} {1166} (\bibinfo {year} {2015})}\BibitemShut
  {NoStop}%
\bibitem [{\citenamefont {Zeh}(2015)}]{Zeh:2015ab}%
  \BibitemOpen
  \bibfield  {author} {\bibinfo {author} {\bibfnamefont {H.~D.}\ \bibnamefont
  {Zeh}},\ }\href {http://arxiv.org/abs/1510.02239} {\bibfield  {journal}
  {\bibinfo  {journal} {arXiv}\ } (\bibinfo {year} {2015})},\ \Eprint
  {http://arxiv.org/abs/1510.02239} {1510.02239} \BibitemShut {NoStop}%
\bibitem [{\citenamefont {Pikovski}\ \emph
  {et~al.}(2015{\natexlab{b}})\citenamefont {Pikovski}, \citenamefont {Zych},
  \citenamefont {Costa},\ and\ \citenamefont {Brukner}}]{Pikovski:2015ab}%
  \BibitemOpen
  \bibfield  {author} {\bibinfo {author} {\bibfnamefont {I.}~\bibnamefont
  {Pikovski}}, \bibinfo {author} {\bibfnamefont {M.}~\bibnamefont {Zych}},
  \bibinfo {author} {\bibfnamefont {F.}~\bibnamefont {Costa}}, \ and\ \bibinfo
  {author} {\bibfnamefont {{\v C}.}~\bibnamefont {Brukner}},\ }\href
  {http://arxiv.org/abs/1509.07767} {\bibfield  {journal} {\bibinfo  {journal}
  {arXiv}\ } (\bibinfo {year} {2015}{\natexlab{b}})},\ \Eprint
  {http://arxiv.org/abs/1509.07767} {1509.07767} \BibitemShut {NoStop}%
\bibitem [{\citenamefont {Pikovski}\ \emph
  {et~al.}(2015{\natexlab{c}})\citenamefont {Pikovski}, \citenamefont {Zych},
  \citenamefont {Costa},\ and\ \citenamefont {Brukner}}]{Pikovski:2015ac}%
  \BibitemOpen
  \bibfield  {author} {\bibinfo {author} {\bibfnamefont {I.}~\bibnamefont
  {Pikovski}}, \bibinfo {author} {\bibfnamefont {M.}~\bibnamefont {Zych}},
  \bibinfo {author} {\bibfnamefont {F.}~\bibnamefont {Costa}}, \ and\ \bibinfo
  {author} {\bibfnamefont {C.}~\bibnamefont {Brukner}},\ }\href
  {http://arxiv.org/abs/1508.03296} {\bibfield  {journal} {\bibinfo  {journal}
  {arXiv}\ } (\bibinfo {year} {2015}{\natexlab{c}})},\ \Eprint
  {http://arxiv.org/abs/1508.03296} {1508.03296} \BibitemShut {NoStop}%
\bibitem [{\citenamefont {Di{\'o}si}(2015)}]{Diosi:2015aa}%
  \BibitemOpen
  \bibfield  {author} {\bibinfo {author} {\bibfnamefont {L.}~\bibnamefont
  {Di{\'o}si}},\ }\href {http://arxiv.org/abs/1507.05828} {\bibfield  {journal}
  {\bibinfo  {journal} {arXiv}\ } (\bibinfo {year} {2015})},\ \Eprint
  {http://arxiv.org/abs/1507.05828} {1507.05828} \BibitemShut {NoStop}%
\bibitem [{\citenamefont {Margalit}\ \emph {et~al.}(2015)\citenamefont
  {Margalit}, \citenamefont {Zhou}, \citenamefont {Machluf}, \citenamefont
  {Rohrlich}, \citenamefont {Japha},\ and\ \citenamefont
  {Folman}}]{Margalit:2015aa}%
  \BibitemOpen
  \bibfield  {author} {\bibinfo {author} {\bibfnamefont {Y.}~\bibnamefont
  {Margalit}}, \bibinfo {author} {\bibfnamefont {Z.}~\bibnamefont {Zhou}},
  \bibinfo {author} {\bibfnamefont {S.}~\bibnamefont {Machluf}}, \bibinfo
  {author} {\bibfnamefont {D.}~\bibnamefont {Rohrlich}}, \bibinfo {author}
  {\bibfnamefont {Y.}~\bibnamefont {Japha}}, \ and\ \bibinfo {author}
  {\bibfnamefont {R.}~\bibnamefont {Folman}},\ }\href
  {http://science.sciencemag.org/content/349/6253/1205} {\bibfield  {journal}
  {\bibinfo  {journal} {Science}\ }\textbf {\bibinfo {volume} {349}},\ \bibinfo
  {pages} {1205} (\bibinfo {year} {2015})}\BibitemShut {NoStop}%
\bibitem [{\citenamefont {Bonder}\ \emph {et~al.}(2015)\citenamefont {Bonder},
  \citenamefont {Okon},\ and\ \citenamefont {Sudarsky}}]{Bonder:2015ab}%
  \BibitemOpen
  \bibfield  {author} {\bibinfo {author} {\bibfnamefont {Y.}~\bibnamefont
  {Bonder}}, \bibinfo {author} {\bibfnamefont {E.}~\bibnamefont {Okon}}, \ and\
  \bibinfo {author} {\bibfnamefont {D.}~\bibnamefont {Sudarsky}},\ }\href
  {http://link.aps.org/doi/10.1103/PhysRevD.92.124050} {\bibfield  {journal}
  {\bibinfo  {journal} {Phys.~Rev.~D}\ }\textbf {\bibinfo {volume} {92}},\
  \bibinfo {pages} {124050} (\bibinfo {year} {2015})}\BibitemShut {NoStop}%
\bibitem [{\citenamefont {Adler}\ and\ \citenamefont
  {Bassi}(2016)}]{Adler:2016aa}%
  \BibitemOpen
  \bibfield  {author} {\bibinfo {author} {\bibfnamefont {S.~L.}\ \bibnamefont
  {Adler}}\ and\ \bibinfo {author} {\bibfnamefont {A.}~\bibnamefont {Bassi}},\
  }\href {http://dx.doi.org/10.1016/j.physleta.2015.10.064} {\bibfield
  {journal} {\bibinfo  {journal} {Phys.~Lett.~A}\ }\textbf {\bibinfo {volume}
  {380}},\ \bibinfo {pages} {390 } (\bibinfo {year} {2016})}\BibitemShut
  {NoStop}%
\bibitem [{\citenamefont {Pang}\ \emph {et~al.}(2016)\citenamefont {Pang},
  \citenamefont {Khalili},\ and\ \citenamefont {Chen}}]{Pang:2016aa}%
  \BibitemOpen
  \bibfield  {author} {\bibinfo {author} {\bibfnamefont {B.}~\bibnamefont
  {Pang}}, \bibinfo {author} {\bibfnamefont {F.~Y.}\ \bibnamefont {Khalili}}, \
  and\ \bibinfo {author} {\bibfnamefont {Y.}~\bibnamefont {Chen}},\ }\href
  {http://arxiv.org/abs/1603.01984} {\bibfield  {journal} {\bibinfo  {journal}
  {arXiv}\ } (\bibinfo {year} {2016})},\ \Eprint
  {http://arxiv.org/abs/1603.01984} {1603.01984} \!\!\!\!\BibitemShut {NoStop}%
\bibitem [{\citenamefont {Pikovski}\ \emph {et~al.}(2016)\citenamefont
  {Pikovski}, \citenamefont {Zych}, \citenamefont {Costa},\ and\ \citenamefont
  {Brukner}}]{Pikovski:2016aa}%
  \BibitemOpen
  \bibfield  {author} {\bibinfo {author} {\bibfnamefont {I.}~\bibnamefont
  {Pikovski}}, \bibinfo {author} {\bibfnamefont {M.}~\bibnamefont {Zych}},
  \bibinfo {author} {\bibfnamefont {F.}~\bibnamefont {Costa}}, \ and\ \bibinfo
  {author} {\bibfnamefont {C.}~\bibnamefont {Brukner}},\ }\href
  {http://dx.doi.org/10.1038/nphys3650} {\bibfield  {journal} {\bibinfo
  {journal} {Nat.~Phys.}\ }\textbf {\bibinfo {volume} {12}},\ \bibinfo {pages}
  {2} (\bibinfo {year} {2016})}\BibitemShut {NoStop}%
\bibitem [{\citenamefont {Bonder}\ \emph {et~al.}(2016)\citenamefont {Bonder},
  \citenamefont {Okon},\ and\ \citenamefont {Sudarsky}}]{Bonder:2016aa}%
  \BibitemOpen
  \bibfield  {author} {\bibinfo {author} {\bibfnamefont {Y.}~\bibnamefont
  {Bonder}}, \bibinfo {author} {\bibfnamefont {E.}~\bibnamefont {Okon}}, \ and\
  \bibinfo {author} {\bibfnamefont {D.}~\bibnamefont {Sudarsky}},\ }\href
  {http://dx.doi.org/10.1038/nphys3573} {\bibfield  {journal} {\bibinfo
  {journal} {Nat.~Phys.}\ }\textbf {\bibinfo {volume} {12}},\ \bibinfo {pages}
  {2} (\bibinfo {year} {2016})}\BibitemShut {NoStop}%
\bibitem [{\citenamefont {M\"uller-Kirsten}(2013)}]{Muller-Kirsten:2013aa}%
  \BibitemOpen
  \bibfield  {author} {\bibinfo {author} {\bibfnamefont {H.~J.~W.}\
  \bibnamefont {M\"uller-Kirsten}},\ }\href@noop {} {\emph {\bibinfo {title}
  {{Basics of Statistical Physics}}}}\ (\bibinfo  {publisher} {World Scientific
  Publishing Company},\ \bibinfo {year} {2013})\BibitemShut {NoStop}%
\bibitem [{\citenamefont {Nash}(1972)}]{Nash:1972aa}%
  \BibitemOpen
  \bibfield  {author} {\bibinfo {author} {\bibfnamefont {L.~K.}\ \bibnamefont
  {Nash}},\ }\href@noop {} {\emph {\bibinfo {title} {Elements of Statistical
  Thermodynamics}}},\ \bibinfo {edition} {2nd}\ ed.\ (\bibinfo  {publisher}
  {Addison-Wesley},\ \bibinfo {year} {1972})\BibitemShut {NoStop}%
\bibitem [{\citenamefont {Mandl}(1988)}]{Mandl:1988aa}%
  \BibitemOpen
  \bibfield  {author} {\bibinfo {author} {\bibfnamefont {F.}~\bibnamefont
  {Mandl}},\ }\href@noop {} {\emph {\bibinfo {title} {{Statistical Physics}}}}\
  (\bibinfo  {publisher} {Wiley Library},\ \bibinfo {year} {1988})\BibitemShut
  {NoStop}%
   \bibitem [{Pri()}]{PrivatePikovski}%
  \BibitemOpen
  \href@noop {} {}\bibinfo {howpublished} {There could be further contributions
  to the decoherence time, which are not captured by the Debye model, e.g. due
  to the size and shape of the system, or to processes, which do not contribute
  to $C_\text{\tiny V}$. Nevertheless the Debye model is expected to give a reliable
  estimate of the effect (private communication with I.~Pikovski).}\BibitemShut
  {Stop}%
\bibitem [{\citenamefont {Schlosshauer}(2007)}]{Schlosshauer:2007aa}%
  \BibitemOpen
  \bibfield  {author} {\bibinfo {author} {\bibfnamefont {M.~A.}\ \bibnamefont
  {Schlosshauer}},\ }\href@noop {} {\emph {\bibinfo {title} {{Decoherence and
  the Quantum-To-Classical Transition}}}},\ \bibinfo {edition} {1st}\ ed.\
  (\bibinfo  {publisher} {Springer-Verlag Berlin Heidelberg},\ \bibinfo {year}
  {2007})\BibitemShut {NoStop}%
\bibitem [{\citenamefont {Breuer}\ and\ \citenamefont
  {Petruccione}(2002)}]{Breuer:2002aa}%
  \BibitemOpen
  \bibfield  {author} {\bibinfo {author} {\bibfnamefont {H.~P.}\ \bibnamefont
  {Breuer}}\ and\ \bibinfo {author} {\bibfnamefont {F.}~\bibnamefont
  {Petruccione}},\ }\href@noop {} {\emph {\bibinfo {title} {{The Theory of Open
  Quantum Systems}}}}\ (\bibinfo  {publisher} {Oxford University Press},\
  \bibinfo {address} {Oxford},\ \bibinfo {year} {2002})\BibitemShut {NoStop}%
\bibitem [{\citenamefont {Romero-Isart}(2011)}]{Romero-Isart:2011aa}%
  \BibitemOpen
  \bibfield  {author} {\bibinfo {author} {\bibfnamefont {O.}~\bibnamefont
  {Romero-Isart}},\ }\href {\doibase 10.1103/PhysRevA.84.052121} {\bibfield
  {journal} {\bibinfo  {journal} {Phys.~Rev.~A}\ }\textbf {\bibinfo {volume}
  {84}},\ \bibinfo {pages} {052121} (\bibinfo {year} {2011})}\BibitemShut
  {NoStop}%
\bibitem [{\citenamefont {Pikovski}(2014)}]{Pikovski:2014aa}%
  \BibitemOpen
  \bibfield  {author} {\bibinfo {author} {\bibfnamefont {I.}~\bibnamefont
  {Pikovski}},\ }\emph {\bibinfo {title} {Macroscopic quantum systems and
  gravitational phenomena}},\ \href@noop {} {Ph.D. thesis},\ \bibinfo  {school}
  {University of Vienna} (\bibinfo {year} {2014})\BibitemShut {NoStop}%
\bibitem [{\citenamefont {Kovachy}\ \emph {et~al.}(2015)\citenamefont
  {Kovachy}, \citenamefont {Asenbaum}, \citenamefont {Overstreet},
  \citenamefont {Donnelly}, \citenamefont {Dickerson}, \citenamefont
  {Sugarbaker}, \citenamefont {Hogan},\ and\ \citenamefont
  {Kasevich}}]{Kovachy:2015aa}%
  \BibitemOpen
  \bibfield  {author} {\bibinfo {author} {\bibfnamefont {T.}~\bibnamefont
  {Kovachy}}, \bibinfo {author} {\bibfnamefont {P.}~\bibnamefont {Asenbaum}},
  \bibinfo {author} {\bibfnamefont {C.}~\bibnamefont {Overstreet}}, \bibinfo
  {author} {\bibfnamefont {C.~A.}\ \bibnamefont {Donnelly}}, \bibinfo {author}
  {\bibfnamefont {S.~M.}\ \bibnamefont {Dickerson}}, \bibinfo {author}
  {\bibfnamefont {A.}~\bibnamefont {Sugarbaker}}, \bibinfo {author}
  {\bibfnamefont {J.~M.}\ \bibnamefont {Hogan}}, \ and\ \bibinfo {author}
  {\bibfnamefont {M.~A.}\ \bibnamefont {Kasevich}},\ }\href
  {http://dx.doi.org/10.1038/nature16155} {\bibfield  {journal} {\bibinfo
  {journal} {Nature}\ }\textbf {\bibinfo {volume} {528}},\ \bibinfo {pages}
  {530} (\bibinfo {year} {2015})}\BibitemShut {NoStop}%
\bibitem [{\citenamefont {Arndt}\ \emph {et~al.}(1999)\citenamefont {Arndt},
  \citenamefont {Nairz}, \citenamefont {Vos-Andreae}, \citenamefont {Keller},
  \citenamefont {van~der Zouw},\ and\ \citenamefont
  {Zeilinger}}]{Arndt:1999aa}%
  \BibitemOpen
  \bibfield  {author} {\bibinfo {author} {\bibfnamefont {M.}~\bibnamefont
  {Arndt}}, \bibinfo {author} {\bibfnamefont {O.}~\bibnamefont {Nairz}},
  \bibinfo {author} {\bibfnamefont {J.}~\bibnamefont {Vos-Andreae}}, \bibinfo
  {author} {\bibfnamefont {C.}~\bibnamefont {Keller}}, \bibinfo {author}
  {\bibfnamefont {G.}~\bibnamefont {van~der Zouw}}, \ and\ \bibinfo {author}
  {\bibfnamefont {A.}~\bibnamefont {Zeilinger}},\ }\href
  {http://dx.doi.org/10.1038/44348} {\bibfield  {journal} {\bibinfo  {journal}
  {Nature}\ }\textbf {\bibinfo {volume} {401}},\ \bibinfo {pages} {680}
  (\bibinfo {year} {1999})}\BibitemShut {NoStop}%
\bibitem [{\citenamefont {Pino}\ \emph {et~al.}(2016)\citenamefont {Pino},
  \citenamefont {Prat-Camps}, \citenamefont {Sinha}, \citenamefont
  {Venkatesh},\ and\ \citenamefont {Romero-Isart}}]{Pino:2016aa}%
  \BibitemOpen
  \bibfield  {author} {\bibinfo {author} {\bibfnamefont {H.}~\bibnamefont
  {Pino}}, \bibinfo {author} {\bibfnamefont {J.}~\bibnamefont {Prat-Camps}},
  \bibinfo {author} {\bibfnamefont {K.}~\bibnamefont {Sinha}}, \bibinfo
  {author} {\bibfnamefont {B.~P.}\ \bibnamefont {Venkatesh}}, \ and\ \bibinfo
  {author} {\bibfnamefont {O.}~\bibnamefont {Romero-Isart}},\ }\href
  {http://arxiv.org/abs/1603.01553} {\bibfield  {journal} {\bibinfo  {journal}
  {arXiv}\ } (\bibinfo {year} {2016})},\ \Eprint
  {http://arxiv.org/abs/1603.01553} {1603.01553} \BibitemShut {NoStop}%
\bibitem [{\citenamefont {Belli}\ \emph {et~al.}(2016)\citenamefont {Belli},
  \citenamefont {Bonsignori}, \citenamefont {D'Auria}, \citenamefont {Fant},
  \citenamefont {Martini}, \citenamefont {Peirone}, \citenamefont {Donadi},\
  and\ \citenamefont {Bassi}}]{Belli:2016aa}%
  \BibitemOpen
  \bibfield  {author} {\bibinfo {author} {\bibfnamefont {S.}~\bibnamefont
  {Belli}}, \bibinfo {author} {\bibfnamefont {R.}~\bibnamefont {Bonsignori}},
  \bibinfo {author} {\bibfnamefont {G.}~\bibnamefont {D'Auria}}, \bibinfo
  {author} {\bibfnamefont {L.}~\bibnamefont {Fant}}, \bibinfo {author}
  {\bibfnamefont {M.}~\bibnamefont {Martini}}, \bibinfo {author} {\bibfnamefont
  {S.}~\bibnamefont {Peirone}}, \bibinfo {author} {\bibfnamefont
  {S.}~\bibnamefont {Donadi}}, \ and\ \bibinfo {author} {\bibfnamefont
  {A.}~\bibnamefont {Bassi}},\ }\href {http://arxiv.org/abs/1601.07927}
  {\bibfield  {journal} {\bibinfo  {journal} {arXiv}\ } (\bibinfo {year}
  {2016})},\ \Eprint {http://arxiv.org/abs/1601.07927} {1601.07927}\!\!\!
  \BibitemShut {NoStop}%
\bibitem [{\citenamefont {Schnabel}(2015)}]{Schnabel:2015aa}%
  \BibitemOpen
  \bibfield  {author} {\bibinfo {author} {\bibfnamefont {R.}~\bibnamefont
  {Schnabel}},\ }\href {http://link.aps.org/doi/10.1103/PhysRevA.92.012126}
  {\bibfield  {journal} {\bibinfo  {journal} {Phys.~Rev.~A}\ }\textbf {\bibinfo
  {volume} {92}},\ \bibinfo {pages} {012126} (\bibinfo {year}
  {2015})}\BibitemShut {NoStop}%
\bibitem [{\citenamefont {Braginsky}\ \emph {et~al.}(1987)\citenamefont
  {Braginsky}, \citenamefont {Ilchenko},\ and\ \citenamefont
  {Bagdassarov}}]{Braginsky:1987aa}%
  \BibitemOpen
  \bibfield  {author} {\bibinfo {author} {\bibfnamefont {V.~B.}\ \bibnamefont
  {Braginsky}}, \bibinfo {author} {\bibfnamefont {V.~S.}\ \bibnamefont
  {Ilchenko}}, \ and\ \bibinfo {author} {\bibfnamefont {K.~S.}\ \bibnamefont
  {Bagdassarov}},\ }\href {http://dx.doi.org/10.1016/0375-9601(87)90676-1}
  {\bibfield  {journal} {\bibinfo  {journal} {Phys.~Lett.~A}\ }\textbf
  {\bibinfo {volume} {120}},\ \bibinfo {pages} {300 } (\bibinfo {year}
  {1987})}\BibitemShut {NoStop}%
\bibitem [{\citenamefont {Lamb}(1996)}]{Lamb:1996aa}%
  \BibitemOpen
  \bibfield  {author} {\bibinfo {author} {\bibfnamefont {J.~W.}\ \bibnamefont
  {Lamb}},\ }\href {\doibase 10.1007/BF02069487} {\bibfield  {journal}
  {\bibinfo  {journal} {Int.~J.~Infr.~Mill.~W.}\ }\textbf {\bibinfo {volume}
  {17}},\ \bibinfo {pages} {1997} (\bibinfo {year} {1996})}\BibitemShut
  {NoStop}%
\bibitem [{\citenamefont {Gabrielse}\ \emph {et~al.}(1990)\citenamefont
  {Gabrielse}, \citenamefont {Fei}, \citenamefont {Orozco}, \citenamefont
  {Tjoelker}, \citenamefont {Haas}, \citenamefont {Kalinowsky}, \citenamefont
  {Trainor},\ and\ \citenamefont {Kells}}]{Gabrielse:1990aa}%
  \BibitemOpen
  \bibfield  {author} {\bibinfo {author} {\bibfnamefont {G.}~\bibnamefont
  {Gabrielse}}, \bibinfo {author} {\bibfnamefont {X.}~\bibnamefont {Fei}},
  \bibinfo {author} {\bibfnamefont {L.~A.}\ \bibnamefont {Orozco}}, \bibinfo
  {author} {\bibfnamefont {R.~L.}\ \bibnamefont {Tjoelker}}, \bibinfo {author}
  {\bibfnamefont {J.}~\bibnamefont {Haas}}, \bibinfo {author} {\bibfnamefont
  {H.}~\bibnamefont {Kalinowsky}}, \bibinfo {author} {\bibfnamefont {T.~A.}\
  \bibnamefont {Trainor}}, \ and\ \bibinfo {author} {\bibfnamefont
  {W.}~\bibnamefont {Kells}},\ }\href {\doibase 10.1103/PhysRevLett.65.1317}
  {\bibfield  {journal} {\bibinfo  {journal} {Phys.~Rev.~Lett.}\ }\textbf
  {\bibinfo {volume} {65}},\ \bibinfo {pages} {1317} (\bibinfo {year}
  {1990})}\BibitemShut {NoStop}%
\bibitem [{\citenamefont {Gieseler}\ \emph {et~al.}(2012)\citenamefont
  {Gieseler}, \citenamefont {Deutsch}, \citenamefont {Quidant},\ and\
  \citenamefont {Novotny}}]{Gieseler:2012aa}%
  \BibitemOpen
  \bibfield  {author} {\bibinfo {author} {\bibfnamefont {J.}~\bibnamefont
  {Gieseler}}, \bibinfo {author} {\bibfnamefont {B.}~\bibnamefont {Deutsch}},
  \bibinfo {author} {\bibfnamefont {R.}~\bibnamefont {Quidant}}, \ and\
  \bibinfo {author} {\bibfnamefont {L.}~\bibnamefont {Novotny}},\ }\href
  {\doibase 10.1103/PhysRevLett.109.103603} {\bibfield  {journal} {\bibinfo
  {journal} {Phys.~Rev.~Lett.}\ }\textbf {\bibinfo {volume} {109}},\ \bibinfo
  {pages} {103603} (\bibinfo {year} {2012})}\BibitemShut {NoStop}%
\bibitem [{\citenamefont {Eibenberger}\ \emph {et~al.}(2013)\citenamefont
  {Eibenberger}, \citenamefont {Gerlich}, \citenamefont {Arndt}, \citenamefont
  {Mayor},\ and\ \citenamefont {Tuxen}}]{Eibenberger:2013aa}%
  \BibitemOpen
  \bibfield  {author} {\bibinfo {author} {\bibfnamefont {S.}~\bibnamefont
  {Eibenberger}}, \bibinfo {author} {\bibfnamefont {S.}~\bibnamefont
  {Gerlich}}, \bibinfo {author} {\bibfnamefont {M.}~\bibnamefont {Arndt}},
  \bibinfo {author} {\bibfnamefont {M.}~\bibnamefont {Mayor}}, \ and\ \bibinfo
  {author} {\bibfnamefont {J.}~\bibnamefont {Tuxen}},\ }\href
  {http://dx.doi.org/10.1039/C3CP51500A} {\bibfield  {journal} {\bibinfo
  {journal} {Phys.~Chem.~Chem.~Phys.}\ }\textbf {\bibinfo {volume} {15}},\
  \bibinfo {pages} {14696} (\bibinfo {year} {2013})}\BibitemShut {NoStop}%
\bibitem [{\citenamefont {Bateman}\ \emph {et~al.}(2014)\citenamefont
  {Bateman}, \citenamefont {Nimmrichter}, \citenamefont {Hornberger},\ and\
  \citenamefont {Ulbricht}}]{Bateman:2014aa}%
  \BibitemOpen
  \bibfield  {author} {\bibinfo {author} {\bibfnamefont {J.}~\bibnamefont
  {Bateman}}, \bibinfo {author} {\bibfnamefont {S.}~\bibnamefont
  {Nimmrichter}}, \bibinfo {author} {\bibfnamefont {K.}~\bibnamefont
  {Hornberger}}, \ and\ \bibinfo {author} {\bibfnamefont {H.}~\bibnamefont
  {Ulbricht}},\ }\href {http://dx.doi.org/10.1038/ncomms5788} {\bibfield
  {journal} {\bibinfo  {journal} {Nat.~Commun.}\ }\textbf {\bibinfo {volume}
  {5}} (\bibinfo {year} {2014})}\BibitemShut {NoStop}%
\bibitem [{\citenamefont {Bohren}\ and\ \citenamefont
  {Huffman}(2008)}]{Bohren:2008aa}%
  \BibitemOpen
  \bibfield  {author} {\bibinfo {author} {\bibfnamefont {C.~F.}\ \bibnamefont
  {Bohren}}\ and\ \bibinfo {author} {\bibfnamefont {D.~R.}\ \bibnamefont
  {Huffman}},\ }\href@noop {} {\emph {\bibinfo {title} {{Absorption and
  Scattering of Light by Small Particles}}}}\ (\bibinfo  {publisher} {Wiley
  Library},\ \bibinfo {year} {2008})\BibitemShut {NoStop}%
\bibitem [{\citenamefont {Frauendorf}(1995)}]{Frauendorf:1995aa}%
  \BibitemOpen
  \bibfield  {author} {\bibinfo {author} {\bibfnamefont {S.}~\bibnamefont
  {Frauendorf}},\ }\href {http://dx.doi.org/10.1007/BF01437069} {\bibfield
  {journal} {\bibinfo  {journal} {Z.~Phys.~D}\ }\textbf {\bibinfo {volume}
  {35}},\ \bibinfo {pages} {191} (\bibinfo {year} {1995})}\BibitemShut
  {NoStop}%
\bibitem [{\citenamefont {Hansen}\ and\ \citenamefont
  {Campbell}(1998)}]{Hansen:1998aa}%
  \BibitemOpen
  \bibfield  {author} {\bibinfo {author} {\bibfnamefont {K.}~\bibnamefont
  {Hansen}}\ and\ \bibinfo {author} {\bibfnamefont {E.~E.~B.}\ \bibnamefont
  {Campbell}},\ }\href {\doibase 10.1103/PhysRevE.58.5477} {\bibfield
  {journal} {\bibinfo  {journal} {Phys.~Rev.~E}\ }\textbf {\bibinfo {volume}
  {58}},\ \bibinfo {pages} {5477} (\bibinfo {year} {1998})}\BibitemShut
  {NoStop}%
\bibitem [{\citenamefont {Hornberger}\ \emph {et~al.}(2004)\citenamefont
  {Hornberger}, \citenamefont {Sipe},\ and\ \citenamefont
  {Arndt}}]{Hornberger:2004aa}%
  \BibitemOpen
  \bibfield  {author} {\bibinfo {author} {\bibfnamefont {K.}~\bibnamefont
  {Hornberger}}, \bibinfo {author} {\bibfnamefont {J.~E.}\ \bibnamefont
  {Sipe}}, \ and\ \bibinfo {author} {\bibfnamefont {M.}~\bibnamefont {Arndt}},\
  }\href {\doibase 10.1103/PhysRevA.70.053608} {\bibfield  {journal} {\bibinfo
  {journal} {Phys.~Rev.~A}\ }\textbf {\bibinfo {volume} {70}},\ \bibinfo
  {pages} {053608} (\bibinfo {year} {2004})}\BibitemShut {NoStop}%
\end{thebibliography}
\end{document}